# LEP3: a low-cost, high-luminosity Higgs factory

**Michael Koratzinos**[1]
*Geneva*
*E-mail:* `m.koratzinos@cern.ch`

The discovery of a relatively light Higgs opens up the possibility of circular $e^+e^-$ Higgs factories. LEP3 is such a machine with emphasis on low cost, since it re-uses most of the LHC infrastructure, including the tunnel, cryogenics, and the two general-purpose LHC experiments Atlas and CMS, with some modifications. The energy reach of LEP3 is 240GeV in the centre of mass, close to the ZH production maximum. Alternative tunnel diameters and locations are possible, including a Higgs factory housed in the UNK tunnel, UNK-L, and a machine located in a new 80 km tunnel in the Geneva region, TLEP, than can further house a very high energy pp collider. The design merits further consideration and a detailed study should be performed, so that LEP3 can be one more option available to the community for the next step in High Energy Physics.

---

[1] Speaker

# 1. Introduction

The recent discovery [1] [2] of the Higgs boson, the last missing link of the Standard Model (SM), and the realization that it is rather light (with a mass of 125-126GeV) has renewed the interest of circular $e^+e^-$ colliders that can serve as a Higgs factories. Common wisdom that any eventual Higgs factory would need to be a linear accelerator turns out not to be accurate. Circular machines enjoy higher luminosities than those of a linear machine around the centre of mass energy ($E_{CM}$) needed for Higgs production via the Higgstrahlung process $e^+e^- \rightarrow ZH$. Moreover, compared to a linear machine, they have the advantage of multiple interaction points and the technology they are based on is mature and carries little risk. The big disadvantage of circular machines is the energy reach – a machine the size of LEP can only go to 250 GeV $E_{CM}$, within the maximum of the ZH process, but cannot reach 350GeV, where the couplings of the Higgs to the top quark could also directly be measured. An 80 km machine can reach 350 GeV $E_{CM}$, but cannot go much higher than that. Whether the superior energy reach of linear machines or the superior luminosity of circular machines is a decisive factor would critically depend on the physics landscape after the first results of the 13-14 TeV run of the LHC are digested: If no supersymmetric or other states are found, probably the superior performance of the circular machines would be the decisive factor for the next large High Energy Physics project. If on the other hand an exciting new state is found within the reach of a linear collider, linear colliders would be the way forward. Today we simply do not know which type of machine will be best, so all options merit a detailed study so that an informative decision can be taken at around 2018.

In this talk I will concentrate on the least expensive option of such a machine, LEP3 [3] [4], in the LHC tunnel. Such an option enjoys an excellent cost to performance ratio as it will re-use the whole of the LHC infrastructure (tunnel, cryogenics system, etc.) including the two large general-purpose LHC experiments. Other tunnel diameters can (and should) also be considered, and as it turns out performance increases linearly with the tunnel circumference.

# 2. The Higgs discovery

A new state has been discovered and the first indications are that it decays as expected for a Standard Model Higgs [5] [6]. This discovery strongly influences the strategy for future collider projects. After the initial verifications that the newly discovered particle is indeed Standard Model-like, we are fast entering the precision measurement era: The new state needs to be characterized, therefore we need to measure as accurately as possible the Higgs branching ratios and related couplings, the Higgs coupling to the top quark, the Higgs quantum numbers, the Higgs mass, the Higgs boson self couplings, its total decay width, etc; We need to verify the (tree-level) structure of the theory: are there Invisible Higgs decays, Exotic decays, or any deviations from SM through higher-order operators?; we need to evaluate (new physics) loop-induced effects; and finally, we might need to re-examine and measure even more precisely EW parameters to over-constrain the theory.

We need to stress here that LHC discoveries at 13-14 TeV (operation expected to start in 2015) will lead to a broader horizon and will strongly influence the strategy for the future.

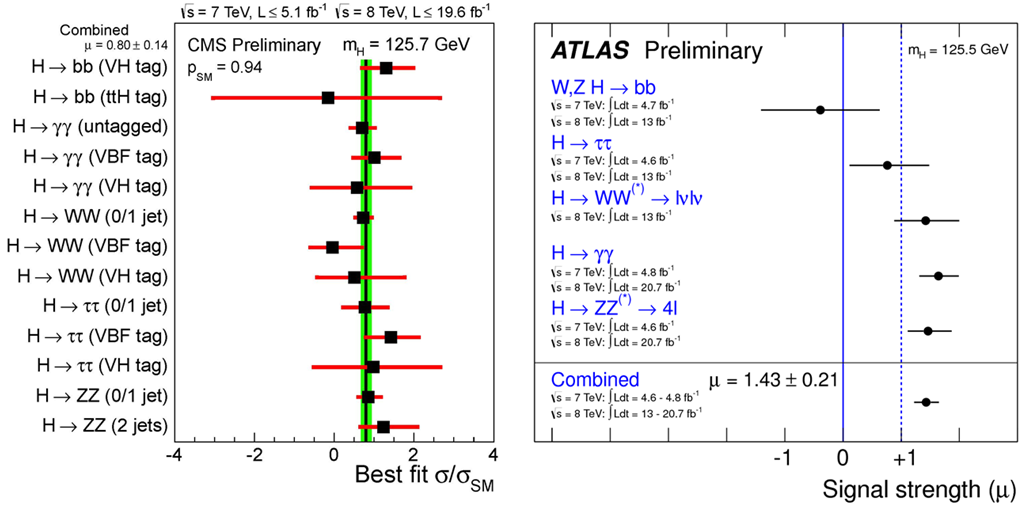

*Figure 1: Higgs decays compared to Standard Model predictions from Atlas and CMS.*

As can be seen from Figure 2, a Higgs with a mass of 125 GeV decays in the most diverse fashion. Many channels are open – so most couplings can be measured from decays, as seen in Table 1 [7]. We should note here the large theoretical uncertainties (of around 5%, mostly QCD) that need to be improved upon before any precision measurements can have an impact. The first experimental results (see Figure 1) do not show any deviations from SM expectations, although errors are large.

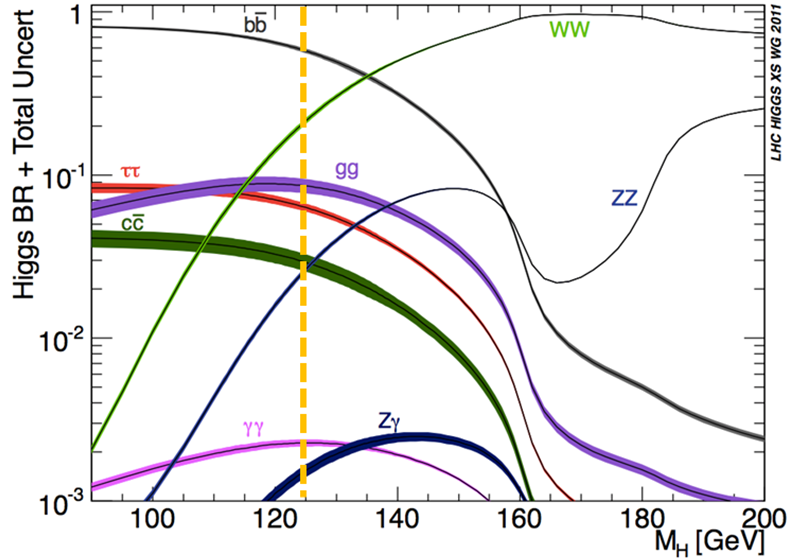

*Figure 2: Higgs brancing fractions as a function of the mass of the Higgs. The orange line reresents the measured mass of the Higgs*

*Table 1: Standard Model Brancing fractions and width of the Higgs and the current theoretical uncertainty.*

| Higgs Decay | BR [%] | Theoretical Uncertainty [%] |
|---|---|---|
| bb | 57.9 | 3. |
| ττ | 6.4 | 6. |
| cc | 2.8 | 12. |
| μμ | 0.022 | 6. |
| WW | 21.6 | 4. |
| gg | 8.2 | 10. |
| ZZ | 2.6 | 4. |
| γγ | 0.27 | 5. |
| Zγ | 0.16 | 9. |
| $\Gamma_H$ [MeV] | 4.0 | 4. |

**3. Precision needed**

The real question after the discovery of the Higgs is what kind of measurement precision is needed [8]. And this is strongly influenced by the expected deviations from Standard Model values predicted by various extensions to the theory. It turns out that for leading contenders of extensions to the Standard Model, the predicted deviations from the SM values are small, typically of order 1%.

Regarding Super Symmetry (SUSY) and more precisely the minimal Supersymmetric Standard Model (MSSM), deviations can be parameterised in terms of two parameters at tree level, the mass of the CP-odd Higgs boson $m_A$ and tanβ. Vector boson couplings are essentially decoupled from EW precision measurements, whereas top and bottom couplings are sensitive to A with mass $m_A$ up to 1TeV. For a moderate tanβ (=5) the typical tree-level coupling modifications due to A are [9]

$$\frac{g_{hVV}}{g_{h_{SM}VV}} \simeq 1 - 0.3\% \left(\frac{200 \text{ GeV}}{m_A}\right)^4$$

(i.e. not very sensitive) whereas

$$\frac{g_{htt}}{g_{h_{SM}tt}} = \frac{g_{hcc}}{g_{h_{SM}cc}} \simeq 1 - 1.7\% \left(\frac{200 \text{ GeV}}{m_A}\right)^2$$

and

$$\frac{g_{hbb}}{g_{h_{SM}bb}} = \frac{g_{h\tau\tau}}{g_{h_{SM}\tau\tau}} \simeq 1 + 1.7\% \left(\frac{1 \text{ TeV}}{m_A}\right)^2$$

are more promising.

Typical coupling modifications from composite Higgs models are also at the per cent level: all couplings reduce together according to the compositeness scale f:

$$\frac{g_{hff}}{g_{h_{SM}ff}} \simeq \frac{g_{hVV}}{g_{h_{SM}VV}} \simeq 1 - 3\% \left(\frac{1 \text{ TeV}}{f}\right)^2$$

In conclusion, expected deviations of leading theories that go beyond the Standard Model from SM predictions are at the per cent level. Any Higgs factory therefore, to be sensitive to new physics, needs to aim to measure to a precision better than this. Since the LHC can also be considered a Higgs factory, the question becomes what is the precision that can be achieved at the LHC? And do we need a different Higgs factory?

## 4. Achievable accuracy of the LHC

Although the LHC is a Higgs factory, and actually creates many more Higgs particles than any $e^+e^-$ Higgs factory can ever hope to create, it is not ideally suited for all Higgs studies: Apart from inherent background considerations when comparing a hadron to a lepton machine, the LHC cannot extract couplings without assumptions on the total width (it can either measure ratios of couplings, or it needs to make assumptions). It is also not sensitive to Higgs invisible decays, something that $e^+e^-$ colliders get for free due to the tagging Z of the Higgstrahlung process, see next section.

The CMS projections on couplings accuracy (under certain assumptions: no exotic decays, no pileup deterioration, stable trigger/detector/analysis performance) can be seen in Table 2 [10]. Atlas numbers are expected to be similar. The sub-percent precision needed for discovery is not available at the LHC, neither for the approved programme (300fb$^{-1}$) or for the high luminosity programme (3000fb$^{-1}$)

*Table 2: CMS-derived uncertainties on LHC measured couplings for 300 and 3000fb$^{-1}$ under different assumptions for the scaling of systematic errors (scenario 1: constant systematic uncertainties; scenario 2: Scaling systematic uncertainties with statistics)*

| Coupling | Uncertainty (%) | | | |
|---|---|---|---|---|
| | 300 fb$^{-1}$ | | 3000 fb$^{-1}$ | |
| | Scenario 1 | Scenario 2 | Scenario 1 | Scenario 2 |
| $\kappa_\gamma$ | 6.5 | 5.1 | 5.4 | 1.5 |
| $\kappa_V$ | 5.7 | 2.7 | 4.5 | 1.0 |
| $\kappa_g$ | 11 | 5.7 | 7.5 | 2.7 |
| $\kappa_b$ | 15 | 6.9 | 11 | 2.7 |
| $\kappa_t$ | 14 | 8.7 | 8.0 | 3.9 |
| $\kappa_\tau$ | 8.5 | 5.1 | 5.4 | 2.0 |

## 5. $e^+e^-$ Higgs factories – choice of beam energy

In an $e^+e^-$ collider the leading Higgs production process is the so-called Higgstrahlung process (see Figure 3 insert), where a Higgs is produced in association with a Z. The presence of an associated Z gives $e^+e^-$ machines a strong advantage, as the invisible width of the Higgs can also be measured. Maximum cross section is at 260 GeV: 212 fb (see Figure 3). If one maximizes physics analysis efficiency (kinematics), luminosity, etc. the most efficient beam energy will actually be smaller than the maximum cross section energy, and in any case at 240 GeV the cross section is only 6% smaller (it now becomes 200 fb), but synchrotron radiation losses of a circular collider are reduced by 40%, therefore 240GeV ECM is the energy of choice for Higgs production in a circular collider.

At this point it should be mentioned that any Higgs factory should also devote part if its beam time in revisiting the Electro Weak precision measurements of LEP, therefore it needs to be able to run at 90 and 160 GeV $E_{CM}$ with good luminosity and possibly transverse polarization for accurate beam energy measurements.

A larger tunnel, like the proposed 80 km tunnel [11] in the Geneva region called TLEP has the added advantage of being able to also run above the top-antitop threshold of 350 GeV.

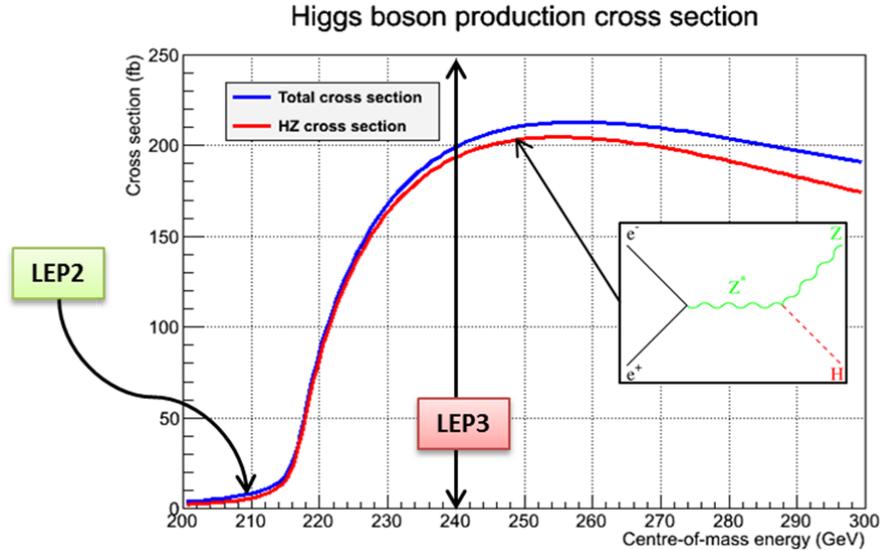

*Figure 3: The Higgs production cross section for the Higgstrahlung process (insert), together with the beam energy of LEP2 and the proposed LEP3 project*

### 6. LEP3 major design considerations

Now that the centre of mass energy is defined, the major design considerations follow the logic underlined below:

- Decide on an acceptable level of SR power dissipation in the ring; in our case, 100MW for both beams. This is four times higher than LEP2.
- As the diameter of tunnel is given, use as high a dipole fill factor as possible. In our case, we have used the readily-available LHeC [12] optics which have a low dipole filling factor (since this was not a priority for the LHeC). An improvement of 10-20% can be expected by moving to a design with a dipole fill factor better than that of LEP2. As it stands, the loss per turn is 7GeV (two times higher than LEP2).
- The above considerations define the total current (7mA). This is again a factor of two higher than LEP2.
- LHeC optics give a 2.5-fold improvement in horizontal emittance, and we use a $\varepsilon_x/\varepsilon_y$ ratio of 500 (LEP achieved 250, but modern light sources achieve numbers in excess of 2000)

- We chose a 1.3GHz (or alternatively a 700MHz) RF system (two or four times higher than LEP2), the small momentum compaction factor giving shorter bunches than LEP2 by a factor of five.
- Chose as small a $\beta^*_y$ value as possible without being too much lower than the longitudinal bunch length (around 3mm). We have chosen a value of $\beta^*_y$ =1mm.
- We should then chose a $\beta^*_x$ value as small as possible but compatible with beamstrahlung limits. We have chosen a value of $\beta^*_x$ =36cm. This can be relaxed if a better horizontal to vertical emittance ratio can be achieved.
- Use as few bunches as possible while keeping within the beam-beam limit. We believe that such a machine is capable of beam-beam tuneshifts of around of 0.09 per IP. The number of bunches satisfying this requirement is four.
- The luminosity we end up with is around $10^{34}\text{cm}^{-2}\text{s}^{-1}$. A sample of the resulting accelerator parameters can be seen in Table 3.

*Table 3: A sample of provisional accelerator parameters for LEP3*

|  | LEP3-H |
| --- | --- |
| $E_{beam}$ [GeV] | 120 |
| circumference [km] | 26.7 |
| beam current [mA] | 7.1 |
| #bunches/beam | 4 |
| #$e-$/beam [$10^{12}$] | 0.98 |
| horizontal emittance [nm] | 25 |
| vertical emittance [nm] | 0.05 |
| bending radius [km] | 2.6 |
| $\kappa_\varepsilon$ | 500 |
| $P_{loss,SR}$/beam [MW] | 50 |
| $\beta^*_x$ [m] | 0.36 |
| $\beta^*_y$ [cm] | 0.1 |
| $\sigma^*_x$ [$\mu$m] | 95 |
| $\sigma^*_y$ [$\mu$m] | 0.22 |
| hourglass $F_{hg}$ | 0.59 |
| $E^{SR}_{loss}$/turn [GeV] | 7.1 |
| $V_{RF}$,tot [GV] | 12 |
| $\xi_x$/IP | 0.07 |
| $\xi_y$/IP | 0.09 |
| $E_{acc}$ [MV/m] | 20 |
| eff. RF length [m] | 600 |
| $f_{RF}$ [MHz] | 700 |
| $\delta^{SR}_{rms}$ [%] | 0.15 |
| $\sigma^{SR}_{z,rms}$ [cm] | 0.31 |
| $\mathcal{L}$/IP[$10^{32}\text{cm}^{-2}\text{s}^{-1}$] | 96 |
| number of IPs | 2 |
| beam lifetime [min] | 16 |

The consequences of such a design with such luminosity is that beams will be "burning up" very fast: the irreducible radiative Bhabha scattering cross section is 0.2 barn, so beam lifetimes will be around 16 minutes for 2 IPs. Therefore, efficient running calls for a "top-up" scheme: a second ring fills the main ring every minute or so, so that the main ring runs at constant energy and with near constant intensity all the time.

Another consequence of high squeezing (to achieve high luminosities) is the appearance of beamstrahlung (see below) which might limit beam lifetimes further.

## 7. Cohabitation issues with the LHC

The big advantage of LEP3 is the re-use of the LHC tunnel and infrastructure, but it poses limits on the timescale of the project. Concurrent operation of LEP3 and the LHC (like in the case of the ring-ring option of LHeC [12]) is not desirable or necessary. LEP3 can fit between the current (14TeV) programme (with or without the still-to-be-approved High Luminosity LHC programme) and the future High Energy LHC programme that necessitates changing all the LHC main magnets. Limited civil engineering works (regarding any bypass tunnels that might be needed) might occur concurrently with the LHC running.

It could well be more economical to leave the main LHC magnets in place while operating LEP3, but in any case ancillary equipment and QRL feed-throughs will have to be dismantled. It has been calculated that the effect of a gentle vertical bend, if the ring is not in the same plane as the IPs, is small [13].

## 8. Beamstrahlung

Beamstrahlung is an important process for circular Higgs factories (as it is for linear ones, but for slightly different reasons) and needs to be discussed in some length: Due to the high focusing of beams at the interaction point, electrons of the one beam see the collective electromagnetic field of the opposite beam and emit photons. This has two effects:

- It alters the $E_{cm}$ of the collision; this is not a problem at LEP3, but is an important consideration at linear colliders where focusing is a lot stronger
- It reduces the beam lifetime, as electrons that fall outside the momentum acceptance of the machine cannot be kept in the ring. This is a problem that is only relevant to circular machines.

The effect of beamstrahlung, which was first pointed out in the context of Higgs factories by V. Telnov [14] is inversely proportional to the horizontal size of the beam at the IP, $\sigma^*_x$, but is independent of $\sigma^*_y$. We have set up a simulation to understand and mitigate the effects of beamstrahlung at LEP3. Mitigation possibilities are large momentum acceptance or very flat beams (or equivalently a high horizontal to vertical emittance ratio). Both of these possibilities do not affect luminosity. On the other hand, reducing the number of electrons per bunch or increasing the longitudinal size of the beam do reduce the beamstrahlung effect, but also reduce luminosity.

Figure 4 shows the photon spectrum of our beamstrahlung simulation, simulated using a detailed collision simulator, Guinea-pig [15]. The effect comes from single hard photon emission, so the effect on lifetime is straightforward: the number of photons with energy higher than the momentum acceptance times the revolution frequency of the machine.

This is on-going work, but early indications show that adequate beam lifetimes can be achieved at LEP3 with emittance ratios of 500 and momentum acceptances around 2%, while sustaining a luminosity of $10^{34}$.

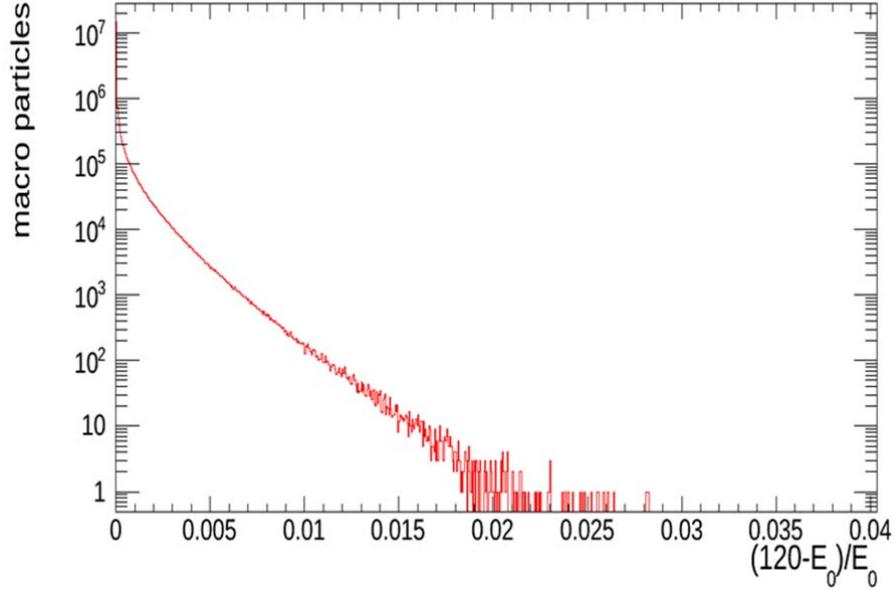

*Figure 4: Photon spectgrum of a typical beamstrahlung simulation. The electrons that have emitted photons with energies above the momentum acceptance of the machine (horizontal scale) are lost, reducing beam lifetimes.*

## 9. RF considerations

As already mentioned, each electron loses 7GeV in a single turn, energy that must be replenished by the RF system (i.e. 7GV would be needed). To actually sustain the large momentum acceptance for mitigating beamstrahlung effects, an additional 5GV of installed RF power might be necessary, bringing the total to 12GV. The top-up ring needs an additional ~7GV, but with low power requirements. Modern superconducting RF cavities can operate with an acceleration gradient of 35MV/m. For LEP3 we will use the much lower value of 20MV/m which can be achieved with better efficiency. Therefore the total length of the RF system turns out to be similar to that of LEP2 (the accelerating gradient is 4 times bigger and the SR losses 4 times higher). Regarding cryogenic power, early calculations show that it would be similar to the currently installed power of the LHC. Geographic location of cryogenic plants (at four IPs around the ring) also matches the envisaged location of RF power. The technology choice of the most appropriate frequency is open: both 1.3GHz and 700MHz have relative merits and drawbacks and the decision will be taken after careful consideration.

## 10. Beam pipe design

Synchrotron radiation issues dominate the beam pipe design which needs to maintain good vacuum and expel the extra SR heat. However, LEP3 is not as demanding as PEPII or SPEAR3 in this respect; the linear power (in units of W/cm) for LEP 3 is 50, compared with 102 for PEPII and 92 for SPEAR. Therefore, the thermal stresses from the SR striking the vacuum chamber should be manageable. On the other hand, the critical photon energy is higher (1.4 MeV) which might lead to radiological risks and activation of material. The beam pipe design will be the subject of a detailed study.

## 11. LEP3 detectors: using the current LHC detectors

An initial study has been performed [16] to assess the suitability of the general-purpose LHC detectors for the physics programme of LEP3. It was found that although the performance was not as good as a dedicated (and especially designed) ILC [17] detector, it nevertheless gave very similar performance, and coupled to the superior luminosity available at LEP3, gave better overall results than the latest ILC study. It must be mentioned that the simulations performed are based on the full simulation of an existing and operational detector, including all its noisy and dead channels, and the description of the detector material tuned with real data. In spite of this, the precision on the Higgs boson couplings studied is significantly better at LEP3, typically a factor 2 to 3, compared to the ILC most recent projections.

## 12. Alternative locations and UNK-L

One obvious alternative location for a LEP3-like accelerator is the UNK tunnel (which I call UNK-L in the following). It has a smaller circumference than the LEP tunnel (but the tunnel itself is wider), so the question is what is the loss in performance compared to LEP3. It turns out that the loss in performance is small. To first order, and for the same SR power dissipation, the luminosity of a circular machine increases linearly with the bending radius. The UNK tunnel is about 30% smaller, but contains less straight sections and the dipole fill factor can be made better than the current LEP3 design. As a result, performance is to within 10% to the proposed LEP3 performance presented here. The loss per turn with a conservative dipole fill factor is 8.3GeV which is higher than LEP3 but manageable. This makes the UNK-L a contender for a possible Higgs factory, provided that funds for the project can be found.

## 13. Comparison with other projects

Although it is not the scope of this paper to compare different approaches to Higgs factories, the question is of relevance for an informed opinion about the relative merits of such projects. The achievable luminosity of LEP3 is slightly better than the one advertised for the ILC [18], for instance, but this is not a decisive factor. The advantage of LEP3 is the low cost and the possibility of multiple interaction points, whereas the advantage of the ILC is the higher energy reach which might or might not be important. Equally, the precision of LEP3 is much better than what can be achieved at the high luminosity LHC for most channels. A larger machine like TLEP would have much higher performance than LEP3, but costs would also be a lot steeper. Regarding the achievable accuracies of different approaches a review can be found in [19].

In conclusion, we cannot answer now the question of if it does make sense to invest in a machine like LEP3 or not. The answer would depend primarily on the physics outcome of the LHC running at 13-14TeV, so we will not know before 2017 or so. If at 2017 the priority would be to measure the Higgs properties, then LEP3 can do it more economically than the ILC and can do it better than HL-LHC. In any case LEP3 remains a good idea that should be investigated further.

*Table 4: relative performance of LEP3 and a possible lepton collider in the UNK tunnel, UNK-L. The proposed LEP3 parameters were used for UNK-L (with the exception of bending radius and circumference)*

|  | **LEP3** | **UNK-L** |
|---|---|---|
| Circumference (m) | 26659 | 20772 |
| Straight sections (m) | 4360 | 3560 |
| Bending radius (physical) | 3549m | 2739m |
| Dipole fill factor | 73% | 80% (LEP2: 87%) |
| Bending radius (km) | 2.6 | 2.19 |
| Eloss/turn (GeV) | 7.06 | 8.35 |
| SR power lost (MW) | 100 | 100 |
| Number of $e^-$ per beam | $4 \times 10^{12}$ | $2.6 \times 10^{12}$ |
| Total beam current (mA per beam) | 7.2 | 6.0 |
| #bunches | 4 | 3 |
| Luminosity (units of $10^{34}$ cm$^{-2}$ s$^{-1}$) | 1.0 | 0.9 |

## 14. Conclusions

The fact that the Higgs is light opens up possibilities for its study that were thought not to be viable – namely circular Higgs factories. LEP3 is a Higgs factory with a modest cost due to re-use of existing infrastructure like the LHC tunnel and cryogenics, and notably the LHC experiments ATLAS and CMS. It is a machine based on proven technology, but pushes accelerator design frontiers in many areas. LEP3 has been conceived only a few months ago and therefore is not at the same state of maturity as other Higgs factory projects. However, initial calculations show that the achievable luminosities ($10^{34}$ cm$^{-2}$s$^{-1}$) would be better than those advertised by the ILC and in addition LEP3 can supply multiple interaction points simultaneously. LEP3 is not extendable to higher energies, so input from the LHC run at 13-14 TeV would be crucial to decide if this is a handicap or not. TLEP is extendable to 350 GeV, plus can house a ~100 TeV proton collider in the future (however its price tag would be similar to that of the ILC). At the moment, one thing is clear: LEP3 (and TLEP) merit further studies.